\documentclass[showpacs,amssymb,floatfix,pre,aps,notitlepage]{revtex4-1}

\usepackage{amssymb,amsfonts,bm}
\usepackage{graphics,graphicx}
\usepackage{color}

\newcommand{\Ltot}{L_{\text{tot}}}
\newcommand{\Lel}{L_{\text{el}}}
\newcommand{\LOSF}{L_{\text{OSF}}}
\newcommand{\xieff}{\xi_{\text{eff}}}

\newcommand{\emma}[1]{#1}

\begin{document}

\title{Bending stiff charged polymers: the electrostatic persistence length}

\author{Emmanuel Trizac}
\affiliation{LPTMS, CNRS, Univ. Paris-Sud, Universit\'e Paris-Saclay, 91405 Orsay, France}
\author{Tongye Shen}
\affiliation{Department of Biochemistry, Cellular and Molecular Biology, 
University of Tennessee, Knoxville, Tennessee 37996, USA}

\begin{abstract}
Many charged polymers, including nucleic acids, are locally stiff.
Their bending rigidity -- quantified by the persistence length --,
depends crucially on Coulombic features, such as the ionic strength of the solution which offers
a convenient experimental route for tuning the rigidity. 
While the classic Odijk-Skolnick-Fixman treatment fails for realistic parameter values,	
we derive a simple analytical formula for the electrostatic persistence length.
It is shown to be in remarkable agreement with numerically obtained Poisson-Boltzmann 
theory results, thereby fully accounting for non-linearities, among which counter-ion condensation effects.
Specified to double-stranded DNA, our work reveals that the widely used bare persistence length 
of 500\,\AA\ is overestimated by some 20\%.
\end{abstract}


\date{\today} 

\maketitle

Since the elucidation of DNA structure in the 1950s, it became 
increasingly clear that the mechanical behavior of a wealth of charged 
polymers (polyelectrolytes) is essential for their interactions
with proteins, the operation of molecular motors and more
generally, their biological function including gene regulation and cytokinesis
\cite{MiMe11,DrLa15}. Fostered in particular by scattering methods and the
more recent advent of single molecule techniques, the 
experimental study of polyelectrolyte
rigidity has consequently been an active field of research
in the last 30 years, see e.g.  
\cite{Tric81,BaJo93,MaSi95,BSBB97,TPSW97,BWAS99,SPEC02,WWRB02,BuBS03,SDCD03,NeOr03,OuMu05,Mann06,Tkac06,SLNP07,TDVK07,BRBB08,SMPR09,VBGT10,KiSu11,NoGo12,BTSR15} 
and references therein. 
It is however arguably one of the most controversial
domain of polymer physics 
\cite{BaJo96,Joan01,EvMY02,NgSh02,Ulln03,NeAn03,DoRu05,Rubin10,Mazu13}. 

Although their mechanical properties may depend on local structure,
polyelectrolytes can satisfactorily be envisioned as coarse-grained
``worm-like'' chains in a variety of situations,
\emma{with a continuous rather than discrete charge distribution}
\cite{MaSi95,BWAS99,BuBS03,SDCD03,SLNP07,ToTh10}. Their flexibility is then 
quantified by a single quantity, the persistence length $\Ltot$,
which measures the distance over which the chain local orientation
decorrelates \cite{GrKh94}. For double-stranded DNA (ds-DNA) in physiological
conditions, $\Ltot \simeq 500\,$\AA, which significantly 
exceeds the typical thickness of the corresponding worm,
having radius $a\simeq 10\,$\AA\ \cite{Hager88,MaSi95,SDCD03}. Unlike 
single-stranded DNA where both lengths are comparable, 
ds-DNA is thus a {\em locally} rigid object. A key question then
lies in the persistence length dependence on external 
control parameters, such as the electrolyte content of the solution
(the so-called ionic strength). 

A major breakthrough is due to
Odijk \cite{Odij77} and independently to Skolnick and Fixman \cite{SkFi77},
who realized that for sufficiently rigid polymers,
the persistence length $\Ltot$ accounts for the bending rigidity
of the polyelectrolyte through the sum of the intrinsic persistence
length $L_0$ of the uncharged polymer, and the electrostatic persistence
length $\Lel$: $\Ltot=L_0+\Lel$.
The presence of charges on the backbone stiffens the chain ($\Ltot     > L_0$),
at least within the mean-field picture adopted by Odijk, Skolnick and
Fixman (OSF). This translates into the celebrated relation 
$\Lel = \LOSF = \lambda^2 \ell_B\, \kappa^{-2}/4$
where $\lambda$ is the line charge of the chain in units of the electron charge
$e$, $\kappa^{-1}$ is the Debye length, and $\ell_B$ is the Bjerrum
length \cite{rque19}. It is important to stress that
in addition to mean-field, the OSF result was derived 
under two stringent conditions: {a)} low charge limit 
(low $\lambda$ or more precisely $\lambda \ll 1/\ell_B$,
where a linearized approach, the so-called Debye-H\"uckel approximation, holds) and
{b)} line charge limit (viewing locally the polymer as a cylinder
of radius $a$, this means enforcing the limit $a\to 0$ or more
precisely, $a \ll 1/\kappa$). It is a sobering thought that
conditions a) and b) are often both violated in practice,
and almost never met simultaneously: 
$\kappa a$ is typically of order 1 for DNA in physiological buffer
while $\lambda \ell_B$ exceeds unity for single-stranded and {\it a fortiori}
double-stranded DNA, as well as for a large gamut of synthetic polymers.
It is generally believed that the non-small $\lambda \ell_B$ 
can be remedied accounting
for counter-ion condensation {\it \`a la}\/ Manning-Oosawa \cite{Mann69,Oosa71,rque10},
with the {\it ad-hoc} replacement 
$\lambda \to 1/(q\ell_B)$ \cite{Odij79,Fixm82,NeAn03} for sufficiently charged chains;
OSF formula, with this prescription, is the mainstay and reference
law for an impressively large body of literature.
%central in the current description of polyelectrolytes.
This approach is to a large extent inadequate, and 
it is thus surprising that the limitations a) and b) above 
have received little attention since early works in the 1980s \cite{Fixm82,LeBr82,rque15}:
%, that pointed out to its limitations, but 
$\LOSF$ may underestimate the mean-field electrostatic persistence
length by an order of magnitude or more. The consequences of
the interpretation of experimental measures, and on a more theoretical
level, on the scaling behavior of $\Lel$, are far reaching. 
Deriving a ``dressed'' or ``renormalized'' OSF formula,
i.e. an analytically simple to use and accurate expression 
for the electrostatic persistence length of rigid polyelectrolytes,
and discussing its physical consequences,
is therefore particularly desirable.
%(for nucleic acids but also synthetic polyelectrolytes). 
It is the main goal of the present study.

We shall work at the mean-field level of Poisson-Boltzmann (PB) theory, that is trustworthy for 
monovalent electrolytes ($q=1$), since ionic correlations
effects are then small \cite{Levi02,BKNN05,M09}. In doing so, we should recover
OSF result for small charges and thin chains, but more importantly,
we will obtain a persistence length of relevance for the 
situations encountered in practice.
At variance with a number of works that operate at Debye-H\"uckel (DH) level 
\cite{BaJo93,PoHP00,HaPo00,EvMY02,NgSh02,Ulln03,ArAn03,MaNe04,Dobr05,KiSu11},
our approach is thus nonlinear. This is an essential prerequisite 
for describing most natural or synthetic polyelectrolytes. Indeed, introducing the dimensionless charge
$\xi=q\lambda \ell_B$, a linearized
treatment is, as a rule of thumb, justified for $\xi\ll 1$ while $\xi>1$ for a wealth 
of synthetic and natural chains (RNA, DNA\ldots).
%(e.g. RNA and DNA in their single- or double-stranded forms). 
Aiming at analytical progress, we coarse-grain  unnecessary atomistic
details, to envision our polymer as a uniformly charged cylinder, a worm-like chain
furthermore assumed to be weakly bent (with a large radius of curvature $R$). 
From the expression of $F$, the bending free energy per unit length for this single
macromolecule in an electrolyte sea, the persistence length follows from
$F = \Lel /(2 R^2)$. 
The calculation is performed within the PB framework
where the dimensionless electric potential $\Psi$ obeys Poisson's equation 
$\nabla^2\Psi = \kappa^2 \sinh \Psi$
(we consider a 1:1 electrolyte, with $q=1$-valent co- and counter-ions). 
To this end, we follow Ref. \cite{Fixm82} and expand $\Psi$ in inverse powers of the curvature:
\begin{equation}
\Psi \,=\, \Psi^{(0)} \,+ \, \frac{1}{\kappa R} \Psi^{(1)} \,+\,
\frac{1}{(\kappa R)^2} \Psi^{(2)}.
\end{equation}
The zeroth order potential is the solution for the straight cylinder problem, 
an already nontrivial analytical problem \cite{Stig75,Rama,TrWi97,OSYa,TrTe06}:
\begin{equation}
\frac{1}{\widetilde r} \,\partial_{\widetilde r}\left(\widetilde r \,\partial_{\widetilde r} \Psi^{(0)}
\right) \,=\, \sinh\Psi^{(0)}
\label{eq:order0}
\end{equation}
where $\widetilde r = \kappa r$. The next order obeys 
\begin{equation}
\frac{1}{\widetilde r} \,\partial_{\widetilde r}\left(\widetilde r \,\partial_{\widetilde r} \Psi^{(1)}
\right)  -\frac{1}{\widetilde r^2}\,\Psi^{(1)}
=\, \Psi^{(1)}\,\cosh\Psi^{(0)}
\label{eq:order1}
\end{equation}
and the equation for $\Psi^{(2)}$ requires the knowledge of $\Psi^{(0)}$ and $\Psi^{(1)}$:
\begin{eqnarray}
&&\frac{1}{\widetilde r} \,\partial_{\widetilde r}\left(\widetilde r \,\partial_{\widetilde r} \Psi^{(2)}
\right) +\frac{1}{2}\left( \partial_{\widetilde r} \Psi^{(1)} + \frac{\Psi^{(1)}}{\widetilde r} -
\widetilde r \partial_{\widetilde r} \Psi^{(0)}
\right) 
\nonumber \\
&& ~~~ =  \Psi^{(2)} \cosh\Psi^{(0)} + \frac{(\Psi^{(1)})^2}{4} \sinh\Psi^{(0)} .
\label{eq:order2}
\end{eqnarray}
While all $\Psi$s vanish for $r\to\infty$,
the bare poly-ion charge sets the boundary conditions at contact ($\widetilde r = \widetilde a \equiv \kappa a $) where the
derivatives of $\Psi^{(0)}$, $\Psi^{(1)}$ and $\Psi^{(2)}$ take the respective values
$-2\xi/\widetilde a$, $2\xi$ and $-\xi\,\widetilde a$. 
The present formulation allows for numerical resolution of the coupled equations 
(\ref{eq:order0}), (\ref{eq:order1}) and (\ref{eq:order2}), from which a classic 
charging process yields the free energy $F$, and thus the persistence length:
\begin{equation}
\Lel \,=\, \frac{2}{\ell_B\,\kappa^2}\, \int_0^\xi \Psi^{(2)}(\widetilde a) \, d\xi.
\label{eq:charging}
\end{equation}
The numerical data presented below have been obtained following these steps,
that also prove useful to proceed analytically, as we now discuss. 

\begin{figure}[htb]
\begin{center}
\includegraphics[width=0.4\textwidth,clip]{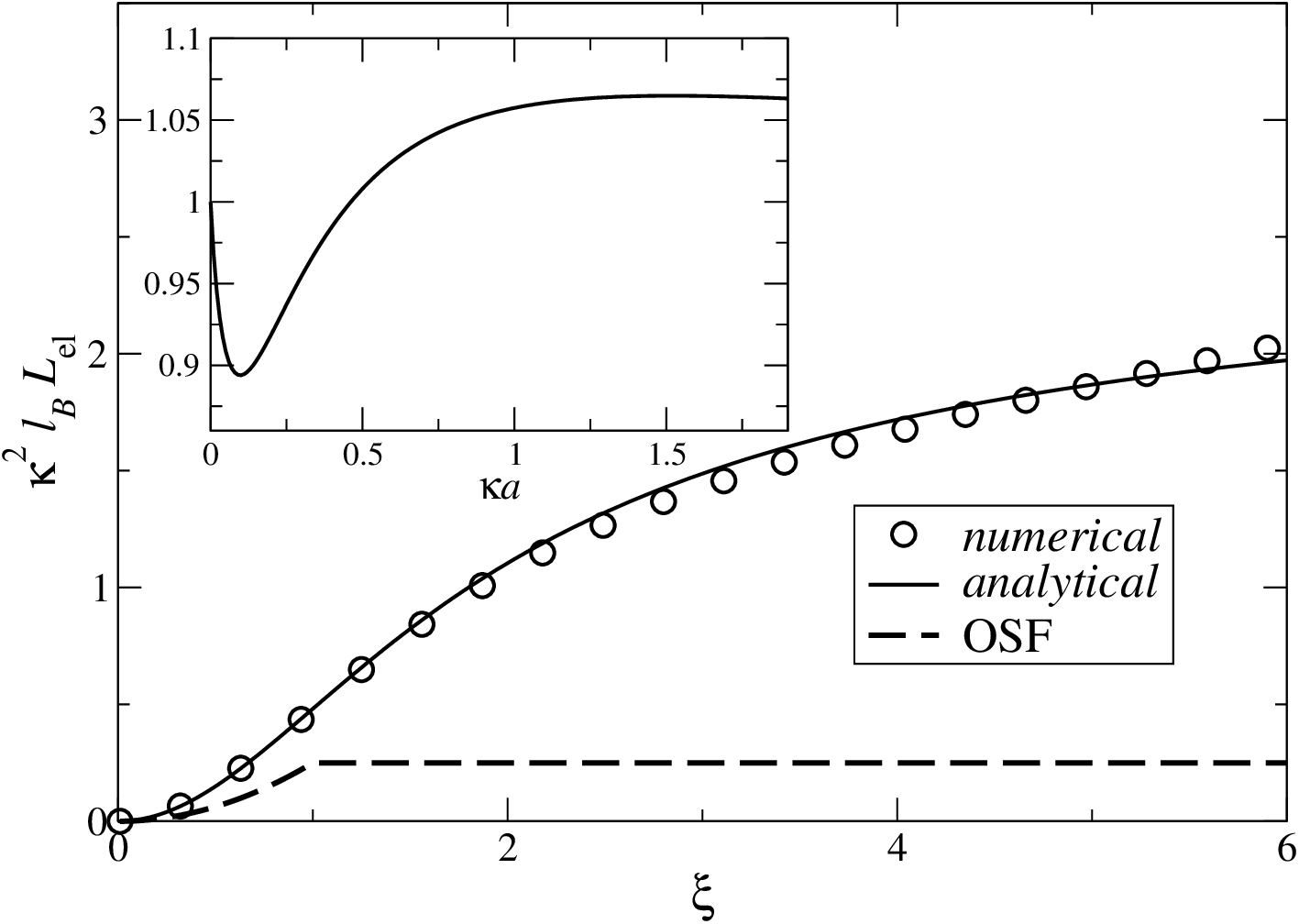}
\caption{Electrostatic persistence length versus bare charge $\xi = \lambda \ell_B$ for
$\kappa a=0.4$. Comparison
of numerical Poisson-Boltzmann results, solving the non-linear coupled 
equations (\ref{eq:order0}), (\ref{eq:order1}) and (\ref{eq:order2}), 
to the analytical surmise Eq. (\ref{eq:Lel}). The OSF law (bottom) is shown by the
dashed line. 
The inset shows the rescaled contact potential $\xi^{-1}\,\Psi^{(2)}_{DH}(\tilde a)/(\tilde a + 1/4)$ 
at linearized (DH) level, remaining close to 1 for all $\kappa a$.}
\label{fig:P_vs_xi}
\end{center}
\end{figure}

Linearizing Eqs. (\ref{eq:order0}), (\ref{eq:order1}) and (\ref{eq:order2}) yields a (DH) description that
should hold for small $\xi$ (i.e. neglecting counter-ion condensation), but valid for arbitrary
$\kappa a$. In other words, deficiency a) above remains while b) is taken care of. 
After tedious calculations, the formula for the contact potential $\Psi^{(2)}(\widetilde a)$ can be 
written explicitly; it turns out immaterial for our purposes,
since a particularly simple linear approximation yields an accuracy better then 10\% for all
$\kappa a$:
\begin{equation}
\Psi^{(2)}_{DH}(\widetilde a) \,\simeq\, \xi \left(\frac{1}{4}+\kappa a\right),
\end{equation}
as shown in the inset of Fig. \ref{fig:P_vs_xi}.
Such a relation is exact for $\kappa a \to 0$ and $\kappa a\to \infty$.
At this linear level of description, the consequence in terms of stiffness is
straightforward: Eq. (\ref{eq:charging})  leads to the Debye-H\"uckel (DH) expression
$\Lel^{DH} = \xi^2 (1+4\kappa a)/(4 \ell_B\kappa^2)$.
For $\kappa a \ll 1$, OSF expression is recovered but the correcting factor $1+4\kappa a$
is in general non-negligible: for instance, considering ds-DNA, having $a\simeq 10\,$\AA,
$\kappa a$ is close to unity in physiological conditions. OSF leads here to a fivefold underestimation.

The remaining task is to account for nonlinearities, that also enhance the Coulombic rigidity compared to OSF expectations.
This requires the analytical resolution of Eqs. (\ref{eq:order0}), (\ref{eq:order1}) and (\ref{eq:order2}),
an intractable task. To circumvent that difficulty, we formulate a surmise that will be gauged later against 
numerics:
\begin{equation}
\Lel \,=\, \frac{\xieff^2}{4 \,\ell_B \,\kappa^2}\, \left(1+4\kappa a\right)
\label{eq:Lel}
\end{equation}
where $\xieff$ is the effective charge of the rod. 
The concept of effective charges is widespread in colloidal science, but it is sometimes elusive.
Here, it has the clear-cut meaning of describing the far-field of the straight charged cylinder \cite{TrTe06,TeTr06},
i.e. $\Psi^{(0)} \sim 2 \xieff K_0(\widetilde r)/[\widetilde a K_1(\widetilde a)]$ for $\widetilde r \gg 1$,
where $K_0$ and $K_1$ denote the 0th and 1st order modified Bessel functions of the second kind.
By construction, $\xieff$ and $\xi$ coincide for weakly charged polymers, while the fact that $\xieff < \xi$
and possibly $\xieff \ll \xi$ for large $\xi$ gives a quantitative meaning to the notion of counter-ion 
condensation. 
Implicit in (\ref{eq:Lel}) is the idea that large-scale features of the
electrostatic interactions dominate for the Coulombic rigidity: 
when bending a straight chain so that it finally has curvature $R$, 
two charges that lie a distance $s$ apart along the backbone become closer by $\Delta s \simeq -s^3/(24 R^2)$,
a rapidly increasing function of $s$.  
In a pairwise (DH) picture for a thin chain, the free energy cost $F$ (and thus $\Lel^{DH}$) is the 
weighted integral of $\Delta s$ times the energy variation $\partial_s (e^{-\kappa s}/s)$. In calculating
that integral, we recover OSF, with the interesting information that large distances mostly do contribute.
This backs up the substitution $\xi\to\xieff$ to account
for nonlinearities, beyond DH, a conclusion also reached in \cite{Tkac06}.
\emma{Similar propositions have been put forward, see e.g. the variational
treatment of Refs. \cite{NeOr03,BTSR15}}.
We can finally invoke progress in theoretical understanding of effective charges made in the last 15 years, 
that provide usable expressions. Here, thick and thin
polymers have to be distinguished, meaning that for small and large $\kappa a$, different expressions should
be used. Specifically, for $\kappa a <1/2$ we took \cite{TrTe06,TeTr06}
\begin{eqnarray}
\xieff &=& 2\kappa a\,K_1(\kappa a)\,\frac{1}{\pi} \, \cosh(\pi \mu) \nonumber \\
\hbox{with }~ \mu &=& \frac{-\pi/2}{\log(\kappa a)+\gamma-\log 8-(\xi-1)^{-1}},
\label{eq:predsmallkap}
\end{eqnarray}
valid, for $\xi>1$ \cite{rque105} and where $\gamma\simeq 0.5772$ is the Euler constant.
For $\kappa a >1/2$, use was made of Eqs. (4) and (5) of Ref. \cite{ATB03}:
\begin{equation}
\emma{\xieff \, =\,  2\, \kappa a \,\,t_\lambda\, +\, 
\frac{1}{2} \,\left( 5-\frac{t_\lambda^4+3}{t_\lambda^2+1}
\right)\, t_\lambda }
\label{eq:predbigkap}
\end{equation}
\emma{where  $t_\lambda = T\left(\xi/(\kappa a + 1/2)\right)$
and the function $T$
is defined as $T(x) = (\sqrt{1+x^2}-1)/x.$}

\begin{figure}[htb]
\begin{center}
\includegraphics[width=0.4\textwidth,clip]{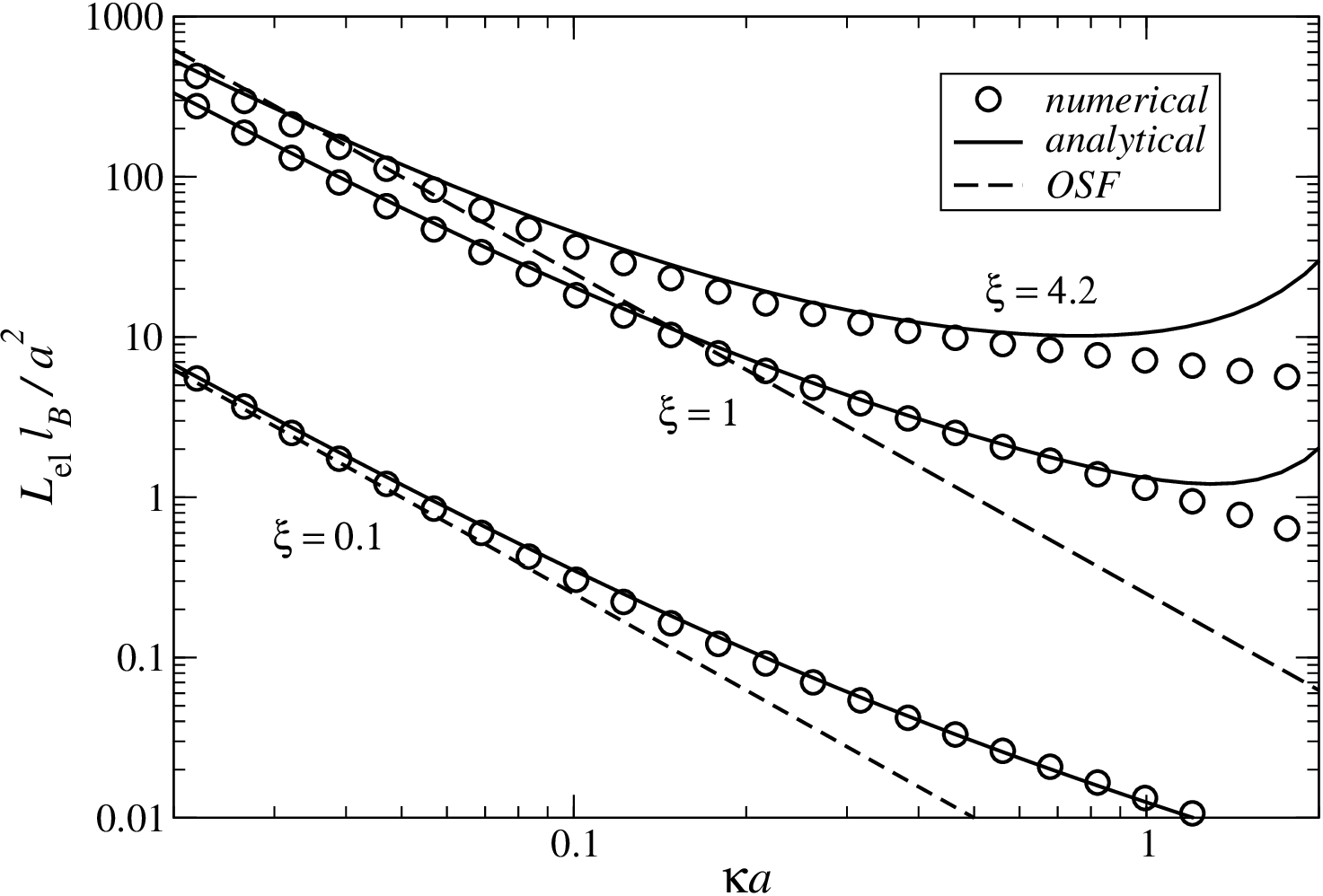}
\caption{Persistence length as a function of salt content, for three charges
(including $\xi=4.2$ for double-stranded DNA). 
OSF (dashed line) does not discriminate between $\xi=1$ and $\xi>1$. The analytical curve is for 
Eq. (\ref{eq:Lel}).}
\label{fig:P_vs_salt}
\end{center}
\end{figure}

We are now in a position to test analytical against numerical results.
First of all, at a particular salt content, Fig. \ref{fig:P_vs_xi} shows that Eq. (\ref{eq:Lel}) 
is remarkably accurate, for all charges.  On the
other hand OSF fails even at small charges, due to the omission of the steric factor $1+4\kappa a$,
and with a growing disagreement as the charge increases. 
Of course, enforcing both $\xi\ll 1$ and $\kappa a \ll 1$, OSF is recovered, as can be seen in 
Figure \ref{fig:P_vs_salt}. This figure also illustrates the quality of the analytical prediction
\emma{for $\kappa a<1$},  
in particular for moderately to strongly charged polymers ($\xi >1$), for which OSF prediction should 
not be employed. \emma{For $\kappa a>1$ the quality of our prediction deteriorates}.

\begin{figure}[htb]
\begin{center}
\includegraphics[width=0.35\textwidth,clip]{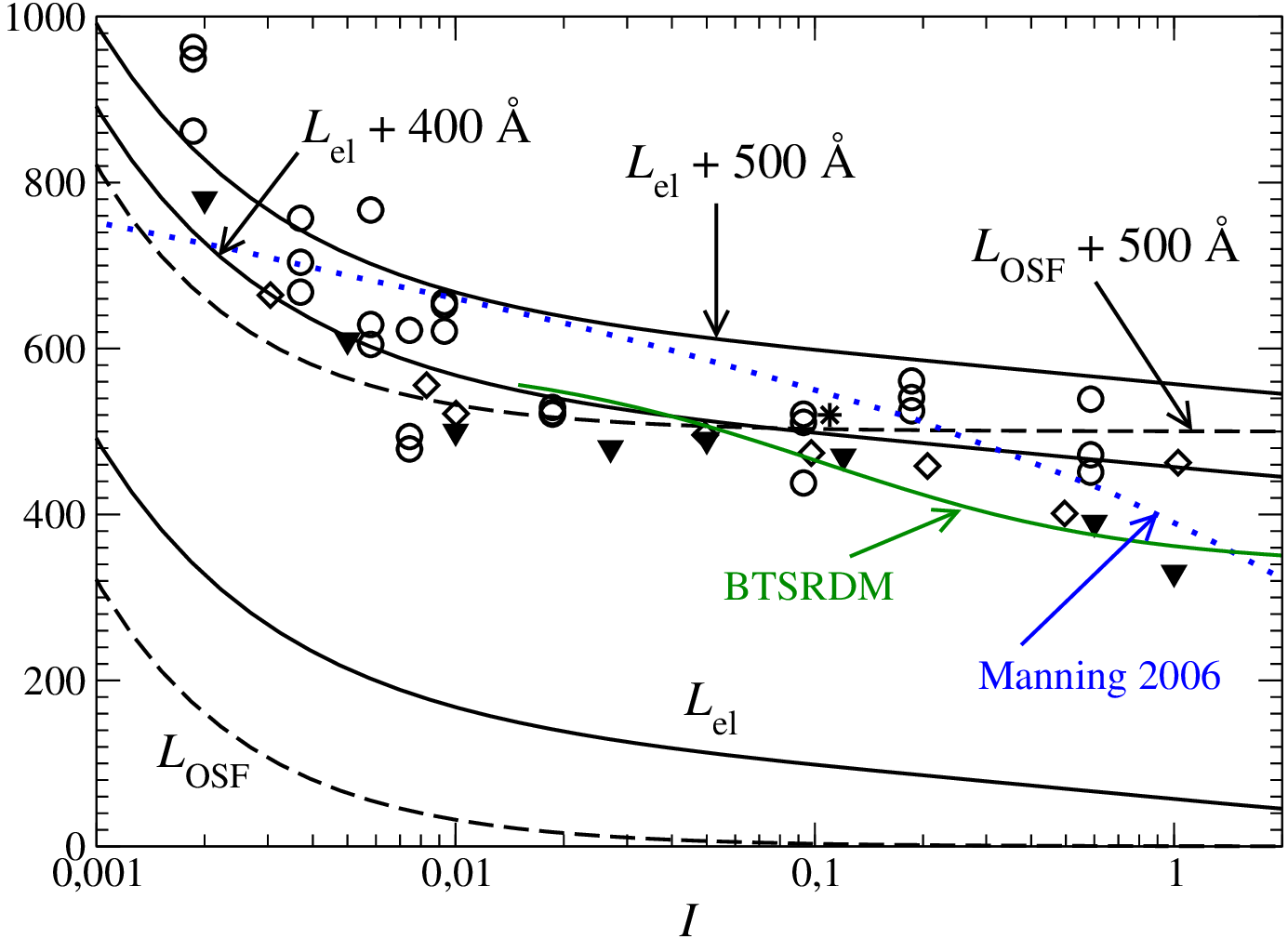}
\includegraphics[width=0.36\textwidth,clip]{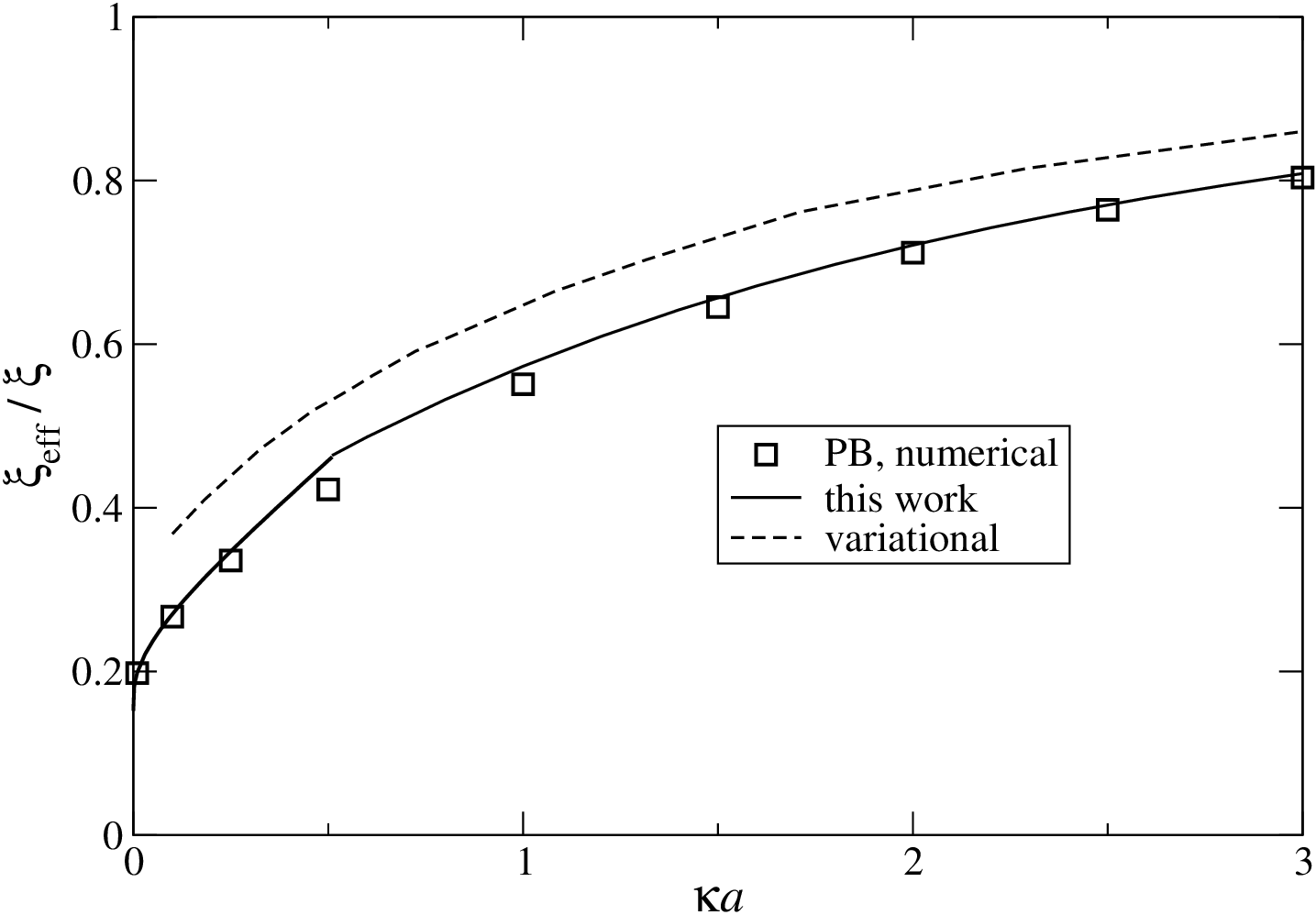}
\caption{Left) Comparison to experimental \emma{and numerical} 
data for ds-DNA: total persistence length in \AA\ as a function
of ionic strength $I$ (salt concentration) in mol\,L$^{-1}$. The latter is defined by 
$\kappa^2 = 8\pi \ell_B I$ so that
$(\kappa a)^2 \simeq 11 \,I$ (with $I$ again in mol\,L$^{-1}$ and $a=10\,$\AA). The diamonds are from \cite{RiSc81} and
the circles show the results from \cite{BSBB97} (three for each ionic strength corresponding
to different methods of calculation). The $\ast$ is for the measure reported in \cite{SDCD03}
($\Ltot=520 \pm 20\,$\AA\ in near physiological conditions).
\emma{The down triangles show the simulation data of \cite{Save12}}.
The bottom part of the graph displays the sole electrostatic
contribution from the present work (continuous line, obtained numerically) and from OSF (dashed line). The bare
contribution $L_0$ is subsequently added, with $L_0=400 \,$\AA\ or $L_0=500 \,$\AA\ 
as indicated. 
\emma{For comparison, the prediction of Ref. \cite{Mann06} is indicated by the 
thick dotted line. Right) Amplitude of charge renormalization effect vs salt
for $\xi=4.2$. The effective charge extracted from the numerical solution of PB equation
(\ref{eq:order0}) is compared to the analytical prediction of equations (\ref{eq:predsmallkap}) and
(\ref{eq:predbigkap}). The result of the variational treatment following Refs.
\cite{NeOr03} and \cite{BTSR15} is also shown (dashed line), together with the 
fit reproducing the 2060 bp experimental data of Ref. \cite{BTSR15} (``BTSRDM'' curve).}}
\label{fig:DNA}
\end{center}
\end{figure}

\begin{figure}[htb]
\begin{center}
\includegraphics[width=0.4\textwidth,clip]{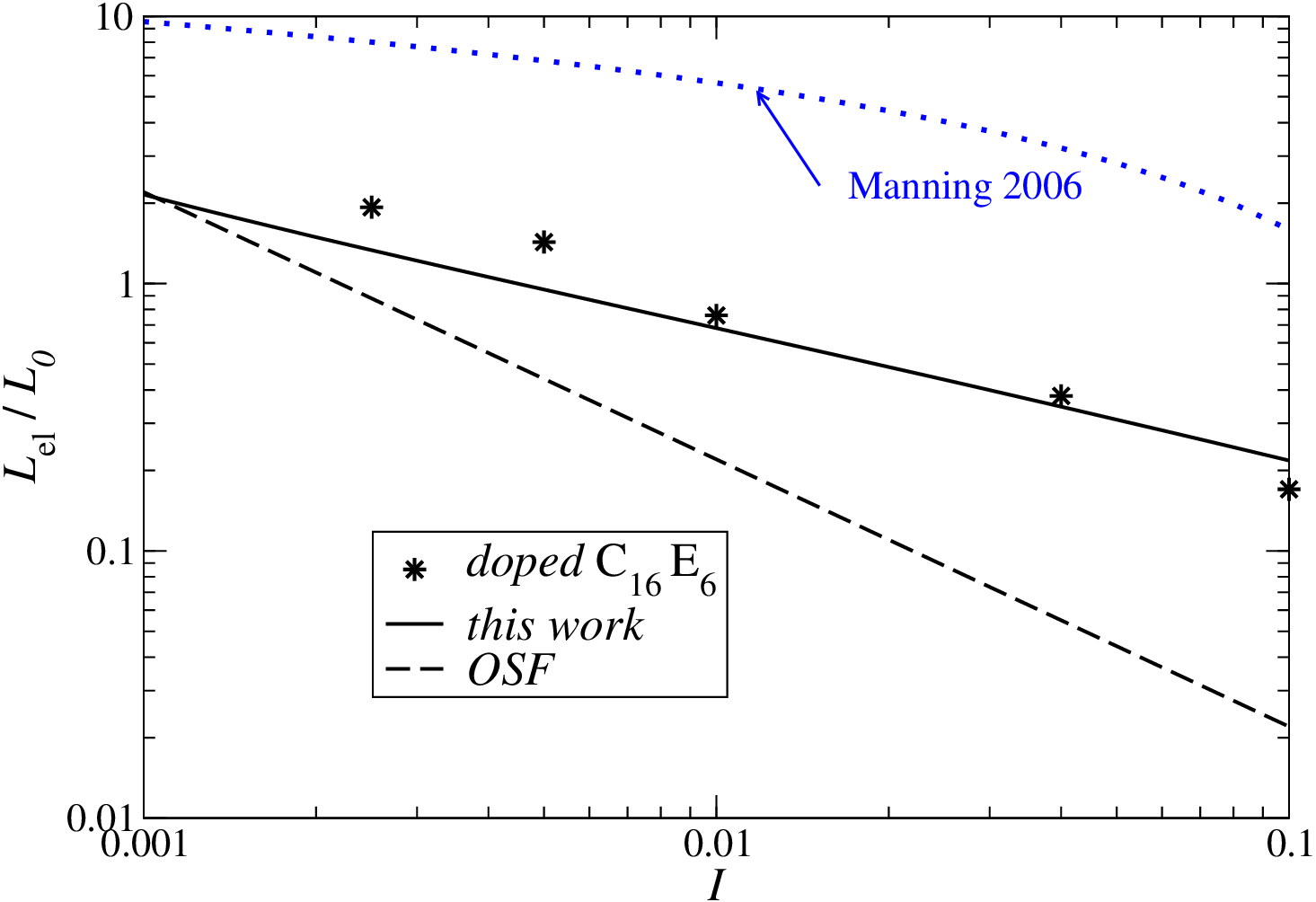}
\caption{Stiffness of worm-like non-ionic micelles (surfactant C$_{16}$E$_6$), doped with ionic 
surfactant to tune the resulting equilibrium polymer charge density \cite{SPEC02,SCEP00}.
The ionic strength $I$ is in mol\,L$^{-1}$.
%(measures by light and neutron scattering).
Here, $\xi=1.2$, $a=30\,$\AA. The bare persistence length $L_0$ can be measured independently:
$L_0=150\,$\AA. }
\label{fig:C16E6}
\end{center}
\end{figure}

Arguably, the most iconic stiff polyelectrolyte is ds-DNA, the mechanical properties of which have been the
subject of a flood of publications. To account for experimental measures, 
its persistence length is almost invariably
fitted assuming OSF, i.e. with the formula $L_0 + \LOSF$ where $L_0$ is unknown. This yields 
the bare length $L_0 \simeq 500\,$\AA, 
a value that is widely taken for granted \cite{Hager88,BSBB97,BlCT00}. 
However, $\LOSF$ underestimates the electrostatic length (see the two bottom curves of Fig.
\ref{fig:DNA}), a deficiency that needs to be compensated by an overestimation 
of $L_0$. The OSF-based value $L_0 \simeq 500\,$\AA\ should thus be reconsidered. 
Fig. \ref{fig:DNA} shows that $L_0+\Lel$ with $L_0 = 400\,$\AA\ provides an 
equally good fit, if not better, than $\LOSF + 500\,$\AA. On the 
other hand, $\Lel + 500\,$\AA\ yields a poor agreement with the experimental data.
The latter are quite scattered, so that no attempt was made at providing
a more accurate estimation of $L_0$.
We conclude at this point that a consistent treatment of Coulombic effects 
at Poisson-Boltzmann level
leads to a ds-DNA bare persistence length that is some 20\% smaller than reported 
in the literature. \emma{It is worth emphasizing here that a distinct in spirit
approach was proposed by Manning in \cite{Mann06}, accounting for the internal tension 
on DNA caused by phosphate-phosphate repulsion. As a consequence, the persistence length
of the uncharged backbone enters multiplicatively into the formula for $\Lel$, and not
additively as here. These results also are seen in Fig. \ref{fig:DNA} to be in fair agreement
with experimental data. We will see however that the formula of Ref. \cite{Mann06}
fails to reproduce the data of Fig. \ref{fig:C16E6}. Finally, we show for completeness
in Fig.\ref{fig:DNA}-right how our effective charge compares to the variational treatment
of \cite{NeOr03,BTSR15}.}

The spread of experimental points in Fig. \ref{fig:DNA} evidences the fact that inferring persistence
length from force-extension curves or other measures is an indirect and delicate 
task \cite{BTSR15,HsPB13}.
For nucleic acids, the bare length $L_0$ is furthermore unknown (charges cannot be ``switched off''),
and we have seen that fitting this quantity within an improper framework
may conceal theoretical glitch. It is thus of particular interest to consider systems 
where the charge can be tuned, and even made to vanish, a limit where $\Ltot$ and 
$L_0$ coincide. 
This is the case of the doped giant micelles studied in \cite{SCEP00,SPEC02}.
The comparison in Fig. \ref{fig:C16E6} is thus fitting-parameter free. Unlike OSF, our approach fares well against
the experiments. Using Manning's formula \cite{Mann06} leads to an overestimation 
of $\Ltot-L_0$ by a factor close to 5, see the thick dotted curve \cite{rque101}.

Before concluding, we briefly comment on the scaling properties of $\Lel$, 
a question that has not been undisputed.  
Weakly charged chains (for which $\xieff \simeq \xi$ is fixed in Eq. (\ref{eq:Lel}))
exhibit two regimes, $\Lel \propto \kappa^{-2}$ for $\kappa a <1/4$ and 
$\Lel \propto \kappa^{-1}$ for $\kappa a >1/4$. 
However, realistically charged polymers reveal
a much weaker dependence on $\kappa$ \cite{rque199}, except under weak screening ($\kappa a \ll1$),
where the standard $\kappa^{-2}$ form is recovered \cite{rque200}. 
Indeed, the effective charge $\xieff$ is an increasing function of salt
density, to such an extent that the $\kappa$ dependence of $\Lel$ becomes 
small for DNA-like parameters, see the $\xi=4.2$ curve in Fig. \ref{fig:P_vs_salt}.
Upon increasing $\xi$ further, the persistence-curves become flatter and flatter
in Fig. \ref{fig:P_vs_salt} (not shown). It is worth emphasizing that
the $\kappa$ dependence of $\xieff$ is not algebraic, so that non-linearities
wipe out the power-law features present in the linear (DH) treatment.
We also stress that at any rate, the OSF scaling in $\kappa^{-2}$
should never be expected, irrespective of the charge of the polymer, 
for $\kappa a>0.1$ \cite{comment_toulouse}. 
%This is what is done in Boue et al.

{\em Conclusion}. Flexibility is a key property of chain macromolecules. We have accounted for the Coulombic 
contribution to the rigidity of stiff polyelectrolytes by a simple formula, Eq. (\ref{eq:Lel}).
It remedies the shortcomings of the celebrated Odijk, Skolnick and
Fixman law, limited to weak screening (thin rods) and weak charges. 
The resulting renormalized treatment is thus applicable to situations of experimental interest,
as we have discussed. A byproduct of our analysis is that the bare ds-DNA persistence length of 500\,\AA\
has been systematically overestimated, and that a consistent value is quite smaller,
$L_0 \simeq 400\,$\AA. We did not present any comparison with single-stranded DNA, since this chain
is considerably more flexible that its double-stranded form. Yet, our description 
\emma{might} be relevant for single-stranded DNA under sufficient tension \cite{DMZP02}.
Finally, our simple approach clearly bears its own limitations. 
While a relevant starting point, the homogeneous worm-like view, subsuming all elastic features in 
a single quantity, is quite crude. It does not account for heterogeneities (like sequence dependence
for nucleic acids), the possible existence of non-smooth bending through flexible joints \cite{ThPe12},
the importance of end effects \cite{ZaRG03}, the coupling between stretching and bending \cite{PoHP00} 
or the fact that elasticity may be scale-dependent \cite{BaJo93}.
In addition, the Poisson-Boltzmann framework discards from the outset specificity effects. It also dispenses with 
ionic correlations, relevant for multivalent ions, 
and that may lead to a decrease of stiffness \cite{BSBB97,NgRS99,GoKL99,ArAn03,Dobr06}, 
somewhat reminiscent of like-charge colloidal attraction.

%%%%%%%%%%%%%%%%%%%%%%%%%%%%%%%%%%%%%%%%%%%%%%%%%%%%%%%%%%%%%%%%%
%\bibliographystyle{apsrev}
%\bibliography{biblio_persist}

\begin{thebibliography}{10}
\bibitem{rque19}
$\ell_B$ corresponds
to the distance where the electrostatic repulsion between two electrons
equals the thermal energy $kT$, which yields $7.1\,$\AA\ in water at room temperature.
\bibitem{rque10}
The idea is that locally cylindrical polyion create a 
logarithmic electrostatic potential that may be strong
enough to bind oppositely charged microions ($q$-valent counterions),
so that the rod takes the effective line charge $e/(q \ell_B)$
whenever $\lambda>1/(q\ell_B)$.
\bibitem{rque105}
More general expressions can be found in \cite{TrTe06,TeTr06};
we focus here on the most relevant $\xi>1$ regime, beyond the
so-called Manning threshold.
\bibitem{rque199}
A similar conclusion was reached in Ref. \cite{NeOr03}.
\bibitem{rque200}
This is due to the fact that the effective charge $\xieff$ goes to a constant
when $\kappa a \to 0$. This property is geometry sensitive, and cannot 
be transposed to planar or spherical macroions.
\end{thebibliography}

\end{document}